%% file: main.tex
\documentclass[5p]{elsarticle}

\usepackage[utf8]{inputenc}
\usepackage{flushend}
\usepackage{xspace}
\usepackage[inline]{enumitem}
\usepackage{color}
\usepackage[dvipsnames]{xcolor}
\usepackage[normalem]{ulem} 
\usepackage{booktabs}
\usepackage[spaces,hyphens]{url}
\usepackage[colorlinks,allcolors=blue]{hyperref}
\newcommand{\showtables}{1}

\usepackage{ifthen}
\usepackage{amssymb}
\usepackage{adjustbox}
\usepackage{tikz}

\newboolean{showcomments}
\setboolean{showcomments}{true} 
\ifthenelse{\boolean{showcomments}}
  {\newcommand{\nb}[2]{
    \fcolorbox{gray}{yellow}{\bfseries\sffamily\scriptsize#1}
    {\sf\small$\blacktriangleright$\textit{#2}$\blacktriangleleft$}
   }
   
  }
  {\newcommand{\nb}[2]{}
   
  }

\newcommand{\destaque}{}
\newcommand\swurl[1]{\footnote{\url{#1}}}
\newcommand\excerpt[1]{\emph{``#1''}}
\newcommand\topic{\vspace{0.1cm}$\rightarrow$\xspace}

\journal{Information and Software Technology journal}

\begin{document}

\begin{frontmatter}



\title{The Organization of Software Teams \\ in the Quest for Continuous Delivery: \\ A Grounded Theory Approach}


\author[usp,serpro]{Leonardo Leite}
\author[ufpa]{Gustavo Pinto}
\author[usp]{Fabio Kon}
\author[ufabc,usp]{Paulo Meirelles}

\address[usp]{University of São Paulo (USP), Brazil}
\address[serpro]{Federal Service of Data Processing (Serpro), Brazil}
\address[ufpa]{Federal University of Pará (UFPA), Brazil}
\address[ufabc]{Federal University of ABC (UFABC), Brazil}

\begin{abstract}

\textbf{Context:} To accelerate time-to-market and improve customer satisfaction, software-producing organizations have adopted continuous delivery practices, impacting the relations between development and infrastructure professionals. Yet, no substantial literature has substantially tackled how the software industry structures the organization of development and infrastructure teams.

\textbf{Objective:} In this study, we investigate how software-producing organizations structure their development and infrastructure teams, specifically how is the division of labor among these groups and how they interact.

\textbf{Method:} After brainstorming with 7 DevOps experts to better formulate our research and procedures, we collected and analyzed data from 37 semi-structured interviews with IT professionals, following Grounded Theory guidelines.

\textbf{Results:} After a careful analysis, we identified four common organizational structures: (1) siloed departments, (2) classical DevOps, (3) cross-functional teams, and (4) platform teams. We also observed that some companies are transitioning between these structures.

\textbf{Conclusion:} The main contribution of this study is a theory in the form of a taxonomy that organizes the found structures along with their properties. This theory could guide researchers and practitioners to think about how to better structure development and infrastructure professionals in software-producing organizations.

\end{abstract}

\begin{graphicalabstract}
\includegraphics[scale=0.4]{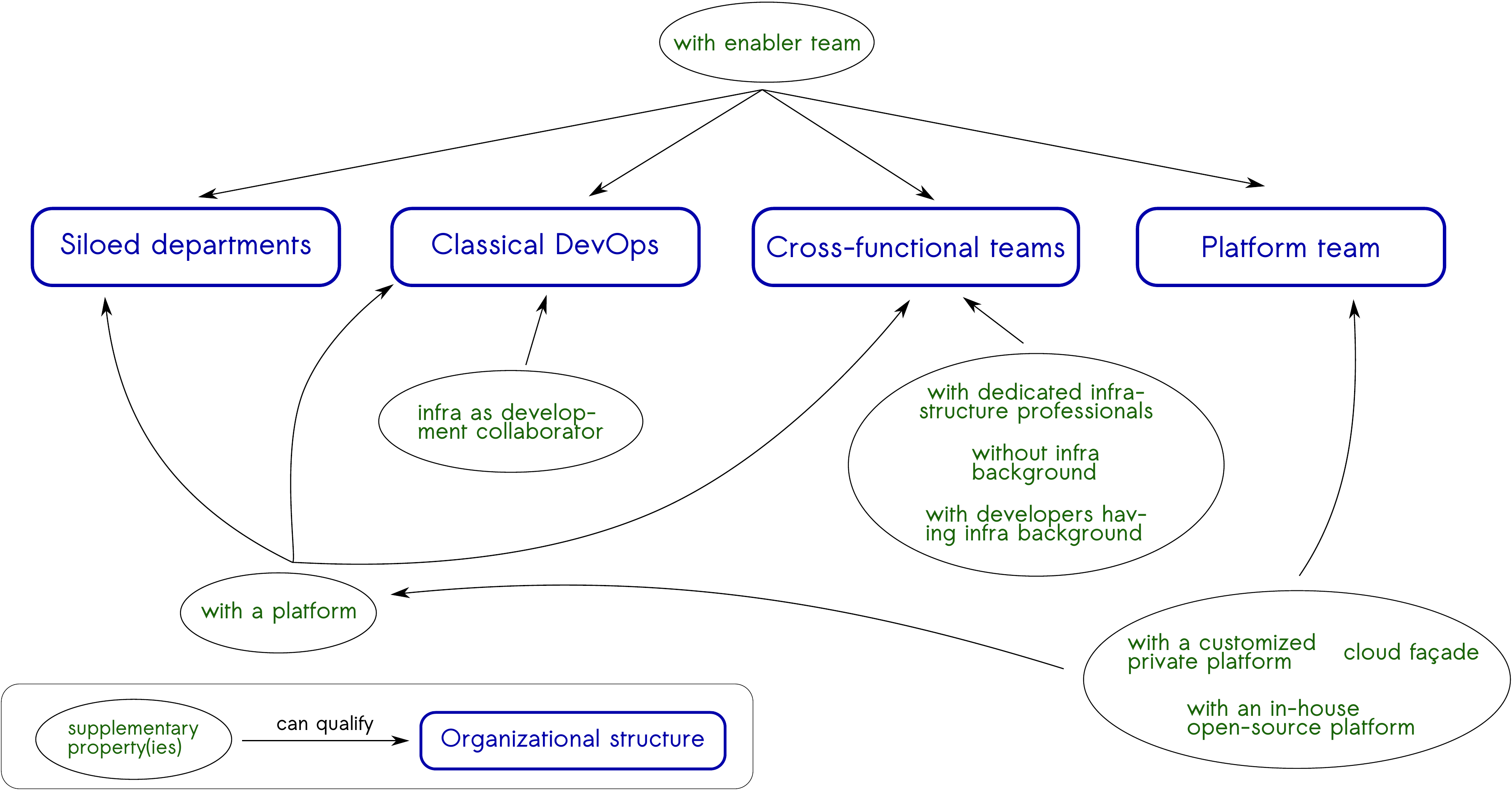}
\end{graphicalabstract}

\begin{highlights}
\item Continuous delivery and DevOps impact organizational structure in software companies
\item We identified four organizational structures for development and infrastructure teams
\item We conceptualized ``platform teams'', an overlooked structure in academic literature
\item Some teams are transitioning across structures, but none to ``siloed departments''
\item The ``platform teams'' structure seems to favor high delivery performance
\end{highlights}


\end{frontmatter}


\input{sections}

\section*{Acknowledgments}

We thank the support of the Brazilian Service of Federal Data Processing (Serpro), CNPq 309032/2019-9, CNPq proc. 465446/2014-0, CAPES – Finance Code 001, and FAPESP proc. 14/50937-1 and 15/24485-9.

\bibliographystyle{elsarticle-num}
\bibliography{bibliography}

\end{document}

%% file: sections.tex

\section{Introduction}

To remain competitive, many software-producing corporations seek to speed up their release processes~\cite{Schermann:2016:ICPC,humble2011devops}. Organizations may adopt continuous delivery practices in their quest to accelerate time-to-market and improve customer satisfaction~\cite{Chen:IEEESW:2015}.
However, continuous delivery also comes with challenges, including profound impacts on various aspects of the software engineering practice~\cite{Schermann:2016:PeerJ}.
With an automated deployment pipeline, one can, for example, question the role of an engineer responsible solely for new deployments. Since release activities involve many divisions of a company (e.g., development, operations, and business), adopting continuous delivery impacts organizational structure~\cite{Chen:IEEESW:2015}.

Given recent transformations, there is a need to better understand the organizational structures that the software industry adopts for development and infrastructure employees\footnote{\destaque In the context of this work, development teams (also called product teams) are responsible for developing business services, while the infrastructure staff uniformly provides computational resources for diverse applications.}. By organizational structure, we mean the differentiation (division of labor) and integration (interaction)~\cite{oliveira2012structures} of operations activities (application deployment, infrastructure setup, and service operation in run-time) among development and operations groups.

However, there is no substantial literature tackling how organizations have structured their development and operations groups. The existing literature presents some classifications for organizational structures~\cite{nybom2016mixing,puppet2018devops,skelton2013topologies,
skelton2019teamtopologies,shahin2017structures}. Still, most of these studies are not based on empirical evidence, which limits the understanding of how the authors conceived their classifications.
{ \destaque 
An exception is the work of Shahin \emph{et al.}~\cite{shahin2017structures}, whose focus was to understand how organizations arrange development and operations teams to embrace continuous delivery practices optimally. However, our quest is not centered around such practices, as it is more general about the structuring of development and infrastructure professionals. Also,  note that social theories are rarely confirmed but are instead corroborated, confronted, or evolved by new studies~\cite{anderson1983cuban,pinfield1986decision,sirmon2011resource}. Our work is unique in the sense that (1) it analyzes a different sample and (2) it follows a different research method; moreover, (3) we discuss the discovered structures in more depth, and (4) we highlight the similarities and differences between the structures discovered by us and Shahin \emph{et al.}, thus bringing more robustness to the knowledge in the area.
}

In this way, this paper addresses the following \textbf{research questions}:

\vspace{0.3cm}
 \fbox{ 
   \parbox{7.5cm}{
     \textbf{RQ1:} \emph{What organizational structures do software-producing organizations adopt to structure operations activities (application deployment, infrastructure setup, and service operation in run-time) among development and infrastructure groups?}

     \textbf{RQ1.1:} \emph{What are the properties of each of these organizational structures?}

     \textbf{RQ1.2:} \emph{Are some organizational structures more conducive to continuous delivery than others?}
     }
 }
\vspace{0.3cm}

To answer these questions, we applied Grounded Theory~\cite{glaser1999gt}, a methodology well-suited for generating theories.
The primary outcome of this research approach is a taxonomy, which is our emerging theory. A taxonomy is a classification system that groups similar instances to increase users' cognitive efficiency, enabling them to reason about classes instead of individual instances~\cite{ralph2019taxonomies}. 

We collected preliminary data in brainstorming conversations with seven specialists, who helped us better understand the relevance of the problem and to shape the questions to be asked in follow-up interviews. We then conducted semi-structured interviews with 37 IT professionals. Based on analysis of the interviews, we discovered four organizational structures:

\vspace{0.3cm}

 \fbox{
   \parbox{7.5cm}{
\textbf{i)} Traditional \emph{siloed departments}, hindering cooperation among development and operations.

\textbf{ii)} \emph{Classical DevOps}, focusing on communication and collaboration among development and operations.

\textbf{iii)} \emph{Cross-functional teams}, taking responsibility for both software development and infrastructure management.

\textbf{iv)} \emph{Platform teams}, providing highly-automated infrastructure services to assist developers. 
     }
 }

\vspace{0.3cm}

For each of these organizational structures, we identified \emph{core} and \emph{supplementary} properties. An organization classified as adopting a given structure will present most of the core properties associated with that structure. Supplementary properties, by contrast, support the explanation of more detailed structural patterns, and an organization may (or may not) exhibit them.

This paper contributes to the area by presenting a systematically-derived taxonomy of organizational structures, based on recent field observations and employing a well-accepted methodology.
In particular, our taxonomy brings the following key benefits: \emph{(i)} it helps practitioners to differentiate \emph{classical DevOps} from \emph{cross-functional teams}, which were traditionally blended under the term DevOps~\cite{humble2011devops,deFranca:2016:SBES}, and \emph{(ii)} it highlights the \emph{platform team} as a promising alternative for organizations.
Moreover, our taxonomy can support practitioners to discuss the current situation of their corporations, supporting decisions on structural changes. It also supports understanding the state of an unknown organization, which can help, for example, engineers in job interviews to evaluate the suitability of working for a given company.

Some of the findings we discuss in this paper relate to \emph{delivery performance}, a construct composed of quantitative metrics, which we explain in Section~\ref{sec:delivery-performance}. We explain our research approach in Section~\ref{sec:research-design}, and we present our taxonomy in detail in Section~\ref{sec:structures}. After this, in Section~\ref{sec:discussion} we discuss the feedback from interviewees and our evaluation. Section~\ref{sec:related-work} discusses related work, whereas Section~\ref{sec:limitations} presents the limitations of this work. Finally, we draw our conclusions and plans for future work in Section~\ref{sec:conclusions}.

\section{Background}
\label{sec:delivery-performance}

Delivery performance combines three metrics: frequency of deployment, time from commit to production, and mean time to recovery~\cite{forsgren2020performance}. It correlates to the organizational capability of achieving both commercial goals (profitability, productivity, and market share) and noncommercial goals (effectiveness, efficiency, and customer satisfaction)~\cite{forsgren2018accelerate}. We used this construct as an indication of how successful an organization has been in adopting continuous delivery. We asked each participant in our study about each of these metrics to define the delivery performance in the interviewee's context.

Based on a survey with 27,000 responses, Forsgren~\emph{et al.}~\cite{forsgren2018accelerate} applied cluster analysis to these metrics and discovered three groups:
\emph{High performers} were characterized as those with multiple deployments per day, commits that take less than 1 hour to reach production, and incidents repaired in less than 1 hour.
\emph{Medium performers} deploying once per week to once per month, had a time from commit to production between one week and one month, and took less than one day to repair incidents.
\emph{Low performers} presented the same characteristics of medium performers for deployment frequency and time from commit to production, but take between one day and one week to repair incidents.

In our research, we are interested in distinguishing between high and non-high performers, not in identifying medium or lower performers. However, the above clusters present a problem, because there is a gap in the values used to identify the high and the medium performers clusters. We circumvented this problem by considering an organization as a high performer if \emph{(i)} it is within the boundaries limiting the cluster of high performers defined above, or \emph{(ii)} it violates no more than one high-performance threshold by no more than one point in the scale adopted for the metric. The scales for each metric are:

\begin{itemize}
\item \emph{Frequency of deployment:} multiple deploys per day;
    between once per day and once per week;
    between once per week and once per month;
    between once per month and once every six months;
    fewer than once every six months.
\item \emph{Time from commit to production:} less than one hour;
    less than one day;
    between one day and one week;
    between one week and one month;
    between one month and six months;
    more than six months.
\item \emph{Mean time to recovery:} less than one hour;
    less than one day;
    between one day and one week;
    between one week and one month;
    between one month and six months;
    more than six months.
\end{itemize}

\section{Study design}
\label{sec:research-design}

This section presents our research approach, including the process we used to collect and analyze data.

\subsection{Grounded Theory}
\label{sec:gt}

Our research aims at generating a theory in the form of a taxonomy for organizational structures in the context of continuous delivery. Broadly speaking, a theory is a system of ideas for explaining a phenomenon~\cite{ralph2019taxonomies}. Taxonomies, on the other hand, are classifications, i.e., collections of classes, wherein each class is an abstraction that describes a set of properties shared by the instances of the class~\cite{ralph2019taxonomies}. If the taxonomy provides explanation, it can be considered a \emph{theory for understanding}: a system of ideas for making sense of what exists or is happening in a domain~\cite{ralph2019taxonomies}. While applying ``taxonomy'' to denote a theoretical formulation, we employ ``classification'' in a broader sense, including classification proposals without theoretical formulation.

Grounded Theory (GT) is a well-suited methodology for generating taxonomies~\cite{ralph2019taxonomies} and a widely used research approach in software engineering~\cite{Stol:ICSE:2016,waterman2015upfront,hoda2017transitions,santos2016jobrotation,deFranca:2016:SBES,luz2019devops}.
A grounded theory must fit the data, predict, explain, have relevance to the field, and be modifiable~\cite{glaser1999gt}.
Its primary advantage is to encourage deep immersion in the data, which may protect researchers from missing instances or oversimplifying and over-rationalizing processes~\cite{ralph2019taxonomies}.
GT is also adequate for our purposes since, according to Stol et al.~\cite{Stol:ICSE:2016}, it is well-suited for questions like ``what's going on here?''. In our case, we want to know what is going on in software-producing organizations that are taking advantage of continuous delivery.
Since multiple GT variants exist, it is important to state which variant we adopt. In this paper, we base our approach on the seminal book \emph{The Discovery of Grounded Theory} from Glaser and Strauss~\cite{glaser1999gt}, which describes what is known as the ``classical Grounded Theory''~\cite{Stol:ICSE:2016}.
While a theory itself must emerge from data, Glaser and Strauss do not disallow pre-established research questions\footnote{Examples of questions guiding sociological inquiries in the GT context: ``Does the incest taboo exist in all societies?'', ``Are almost all nurses woman?''~\cite{glaser1999gt}.}.

The \emph{constant comparative method} is the core method for producing a grounded theory. It relies on rigorous analysis of qualitative data, and it is accomplished with \emph{coding}, a process of condensing original data into a few words with conceptual relevance, which give emergence to theoretical concepts. Glaser and Strauss do not prescribe a precise coding format~\cite{glaser1999gt}, sufficing that the researcher annotates concepts and adheres to the following rules: \emph{(i)} when noting a concept, compare this occurrence with the previous occurrences of the same or similar concepts and \emph{(ii)} while coding, if conflicts and reflections over theoretical notions arise, write a memo on the ideas. A memo is an unstructured note reflecting the researcher's thoughts at a specific point in time. We describe in more detail in Section~\ref{sec:analyzing-interviews} how we applied coding.

Besides the rigorous analysis of qualitative data, GT also relies on the researcher's \emph{theoretical sensitivity}; i.e., their capacity to have theoretical insight into a substantive area. Our theoretical sensitivity comes from direct experience in the IT industry and our previous works on DevOps and software engineering~\cite{siqueira2018gov,cukier2018startup,luz2019devops}, especially a survey on the DevOps literature~\cite{leite2019survey}. This acquired theoretical sensitivity explains our capacity to pose the research questions of this paper.

This article also builds on our recent work, an extended abstract that briefly presents the four organizational structures of our taxonomy~\cite{leite2020building} and a short paper presenting the platform team structure only~\cite{leite2020platform}. In contrast, the current paper describes in detail all four organizational structures and their properties.

In GT, data collection and analysis co-mingle and build on each other, so the emerging theory guides which data to sample next, considering gaps and questions suggested in prior analysis. This process---called \emph{theoretical sampling}---avoids the usual statistical notions of verificational methods, such as significant sample.
Instead, researchers must establish the theoretical purpose of the sample, defining multiple comparison groups, maximizing variation among groups to identify similarities, and minimizing variation to determine differences. We approached theoretical sampling primarily by: \emph{(i)} valuing the diversity of people and organizations in our sample, strengthening the transferability of our theory; and \emph{ii)} interviewing people in contexts in which we could explore hypotheses weakly supported by the chain of evidence built so far. We elaborate more on the choices of participants in Section~\ref{sec:selection}.

Ideally, the researcher continues the analysis until \emph{theoretical saturation} is achieved, which means that new data no longer meaningly impacts the theory. In this work, reaching saturation advances our previous publications~\cite{leite2020platform,leite2020building}. Our criteria for saturation is described in Section~\ref{sec:saturation}. As an ever-evolving entity, rather than a finished product, new data can always be analyzed to alter or expand a grounded theory. Accordingly, practitioners could (and potentially will) adjust the theory when applying it to their concrete scenarios~\cite{glaser1999gt}. Therefore, in this work, we present an \emph{emerging theory}, rather than a fully-validated one, which we explain more in Section~\ref{sec:evaluation}.

We applied the GT techniques to data from interviews with IT professionals. In the next sections, we present how we chose our subjects and the design and analysis of these interviews.

\subsection{Brainstorming sessions}

After drafting our research questions, we conducted ``brainstorming sessions'' with seven specialists experienced with DevOps. Some of them have witnessed DevOps transformations in large organizations, while others have actively shaped such transformations in large and small companies.

The base script for these sessions aimed to elicit feedback on our research questions and spark discussion of concerns raised by our survey on the DevOps literature~\cite{leite2019survey}. The conversations were essential for us to fine-tune our research questions and research approach. These sessions also helped us to better target the script for the following semi-structured interviews toward concerns learned from these experts. {\destaque Therefore, the concrete outputs of this phase were: the research questions present in this text and the interview script (explained in Section~\ref{sec:conducting-interviews}).}

We did not apply the analysis procedures detailed in Section~\ref{sec:analyzing-interviews} for these preliminary conversations, considering they were not intended to provide answers to our research questions. Nevertheless, we did not dismiss the theoretical insights provided by them. For example, the notion of a \emph{platform team} began to take shape in the brainstorming sessions and, thus, influenced subsequent analysis. After these brainstorming sessions, we started the semi-structured interviews.

\subsection{Selecting participants}
\label{sec:selection}

We sent about 90 interview invitations using a convenience approach: the first invitations were close contacts in our research group's network. We also contacted participants suggested by our interviewees and colleagues. The only requirement was that the participant should work in an industrial context that has adopted continuous delivery or is implementing efforts toward it. Some invited participants did not reply, and some demonstrated interest in participating, but could not make time for it. Ultimately, we interviewed 37 IT professionals ($\approx$ 41\%).
Following ethical procedures~\cite{strandberg2019ethical}, all the interviewees and their organizations are anonymized in this paper. We recorded the interviews for later analysis, keeping the audio records under restricted access. We conducted the interviews from April 2019 to May 2020. Nine interviews were conducted in person, five of which took place at the interviewee's company, and 28 were held online. The sessions took 50 minutes on average (minimum of 24 and a maximum of 107 minutes).

We employed several strategies to foster diversity and enhance comparison possibilities in our sample, as recommended by GT guidelines~\cite{Stol:ICSE:2016}. We aimed to include a broad range of organization and interviewee profiles. For instance, we selected organizations of different sizes\footnote{We employed 200 and 1,000 as size thresholds because they are used in information publicly available on LinkedIn.}: small (30\%), medium (32\%), and large (38\%); of different types: private (90\%), governmental (5\%), and others (5\%);
and from different sectors and countries. We included male (73\%) and female (27\%) professionals, and also chose interviewees with different roles.

\ifthenelse{\equal{\showtables}{1}}{
\begin{table}[ht]
  \caption{Number and description of participants and organizations}
  \label{tab:participants}
  \begin{tabular}{l|l}
    \toprule
    \textbf{Role}                   & \textbf{Team location} \\
    17: Developer                   & 21: Brazil \\
    7: Development manager          & 5: USA  \\
    5: External consultant          & 4: Globally distributed \\
    3: Infrastructure manager       & 2: Germany  \\
    2: Infrastructure engineer      & 2: Portugal  \\
    1: Executive manager            & 1: France \\
    1: Enabler team member          & 1: Canada    \\
    1: Designer                     & 1: Italy  \\
    \textbf{Gender}                 & \textbf{Number of employees} \\
    27: Male                        & \textbf{in the organization} \\
    10: Female                      & 14: More than 1000 \\
    \textbf{Time since graduation}  & 12: From 200 to 1000 \\
    15: More than 10 years          & 11: Less than 200 \\
    13: From 5 to 10 years          & \textbf{Organization type} \\
    9: Less than 5 years            & 33: Private for profit  \\
    \textbf{Degree}                 & 2: Governmental \\
    22: Undergraduate               & 1: Private nonprofit \\
    13: Masters                     & 1: International organization \\
    2:  PhD                         &  \\
  \bottomrule
\end{tabular}
\end{table}
}{TABLE}

Table~\ref{tab:participants} describes the participants, presenting only an aggregated profile of participants to cope with anonymization~\cite{strandberg2019ethical,santos2016jobrotation}. Location refers to that of the interviewee's team; we had four participants working remotely for globally distributed teams. When describing roles, enabler team indicates a specialized technical team that supports developers, but without owning any services. For interviews with consultants, the number of employees refers to the size of the companies that contracted the consultants (and not the consultant's employers). The interviewees worked in the following business domains: IoT, finances, defense, public administration, justice, real estate, maps, education, Internet, big data, research, insurance, cloud, games, e-commerce, telecommunication, fashion, international relations, mobility, office automation, software consulting, inventory management, vehicular automation, team management, and support to software development. Five of the interviewed companies are currently considered unicorn startups, and two of them are tech giants.

We also selected participants with theoretical purposes in mind, thus applying theoretical sampling.
We interviewed participants who work in scenarios where it is particularly challenging to achieve continuous delivery (e.g., IoT, games, or defense systems), aiming to understand the limits and eventual corner cases.
After the twentieth interview, we more actively sought people: in a cross-functional team, in (or interacting with) a platform team, with no (or few) automated tests, with monolithic systems, and labeled as ``full-stack engineers''. We adopted these criteria due to hypotheses not well-supported by our chain of evidence at that time.

To support our findings, we included excerpts from these conversations in our chain of evidence (in the accompanying supplementary material) and in this article.
The excerpts are formatted in italics and within quotes. Excerpts and other accounts refer to interviews using tokens in the format ``\#I\emph{N}''. Thus, ``\#I2'' refers to the second interview, interviewee, or interviewee's organization. Brainstorming sessions are indicated as ``\#B\emph{N}''. Such excerpts and citations are intended to make readers ``feel they were also in the field,'' a GT recommendation~\cite{glaser1999gt}.

\subsection{Conducting the interviews}
\label{sec:conducting-interviews}

Since our goal is to discover existing organizational structures, and not to verify a preconceived set of structures in the field, it would be unsuitable to use only closed questions; instead, we conducted semi-structured interviews. Semi-structured interviews mix closed and open-ended questions, often accompanied by ``why'' and ``how'' questions; the interview can deviate from the planned questions, allowing for discussion of unforeseen issues~\cite{adams2010interviews}, which fits the purpose of theory generation. With semi-structured interviews, we could also focus on different topics in different conversations, according to the relevance of each theme for each context.

Before starting the interviews, we built an interview protocol\swurl{http://ccsl.ime.usp.br/devops/2020-06-14/interview-protocol.html} to guide the process based on our previous experience with interviews~\cite{cukier2018startup}, on other relevant works~\cite{hoda2017transitions,adams2010interviews}, and on the guidelines offered by a website for journalists, called ijnet\swurl{http://ijnet.org/en}.

The interview protocol contains the questions that drove the interviews, which mainly were derived from the brainstorming sessions and our survey of the DevOps literature~\cite{leite2019survey}, and hence also grounded in data\footnote{GT also considers the library as a source of data.}.
The themes addressed by the interview questions include: (1) interviewee company and role; (2) responsibility for deployment, building new environments, non-functional requirements, configuring and tracking monitoring, and incident handling, especially after-hours; (3) delivery performance {\destaque (using the metrics and scales defined in Section~\ref{sec:delivery-performance})}; (4) future improvements in the organization; (5) effectiveness of inter-team communication; (6) inter-team alignment for the success of the projects; (7) description of DevOps team or DevOps role, if existing; and (8) the policy for sharing specialized people (e.g., security and database experts) among different teams.

The interview protocol is not a static document. As we conducted interviews, we changed how we asked some questions, focused more on some questions and less on others, and created new questions to explore rising hypotheses. We provide some indications about the evolution of the questions in the interview protocol itself.

\subsection{Analyzing the interviews}
\label{sec:analyzing-interviews}

We followed the core Grounded Theory principles of \emph{constant comparative method} and \emph{coding}, which are intended to discipline the creative generation of theory. During this process, we created two artifacts for each interview: the \textbf{transcripts} and the \textbf{codes}. We also created two global artifacts: the \textbf{comparison sheet} and the \textbf{conceptual framework}. Finally, by analyzing, comparing, and using all these artifacts, we elaborated our \textbf{taxonomy}, which is the theory itself.

\vspace{0.2cm}
\noindent
\emph{\textbf{Transcripts.}}
We listened to each audio record and transcribed it.
We did not transcribe the full interview. Instead, we summarized relevant parts, excluding minor details and meaningless noise~\cite{georgieva2008lens}.
For instance, we transcribed the following part of a conversation:

\vspace{0.2cm}

{
\small
\emph{``If you break the SLA, there are consequences. You have to improve things; you can't go back to feature development until SLA has recovered. Any problem in final service: developer is paged. If it's infrastructure-related, developers call the infrastructure team. And we solve together. We try to help anyway, because at the end of the day if users can't use the system, we all suffer.''}
}

\vspace{0.2cm}
\noindent
\emph{\textbf{Codes.}}
After transcribing an interview, we then derived the coding by condensing the interviewers' transcripts into a few words (the essence of the interview in relation to our research questions). Each interview has its list of coding, representing the particular reality of that interviewee. The above fragment of transcription, for example, led to the following coding:

\begin{quote}
\small
\emph{
Developers $\rightarrow$ own the availability of their services\\
Broken SLA $\rightarrow$ blocks feature development\\
Broken SLA $\rightarrow$ page developers\\
Broken SLA $\rightarrow$ if needed, call infra
}
\end{quote}

The supplementary material ``chain of evidence'' \\ presents more examples of the coding we performed.

\vspace{0.2cm}
\noindent
\emph{\textbf{The comparison sheet.}}
To support the \emph{constant comparison} of different interviews and codings, we summarized the main characteristics of each interview in a spreadsheet. We filled the cells with concise statements, with rows representing interviews and columns including the following characteristics: interview number, organizational structure, supplementary properties, delivery performance, observation, continuous delivery, microservices, cloud, other teams, non-functional requirements (NFRs), monitoring / on-call, alignment, communication, bottleneck, responsibility conflicts, database, security, specialists sharing, and DevOps team/role.

\vspace{0.2cm}
\noindent
\emph{\textbf{Conceptual framework.}}
From the constant comparison of codes of different interviews emerged theoretical concepts. These theoretical concepts and their relations form the unified \emph{conceptual framework}, which constitutes the shared understanding among the authors about the analyzed data~\cite{miles2013qualitative}. Our conceptual framework is not a representation of our theory, but rather an intermediary artifact used to consolidate, in a single place, the concepts yielded by the coding process, working as the source of concepts to the shaping of our theory.

We maintained a visual representation of our conceptual framework as the research evolved. We provide all of its versions in the supplementary materials. Figure~\ref{fig:conceptual-framework-fragment} provides as example a fragment of our conceptual framework. In that figure, rectangles represent \emph{concepts} abstracted from data, while rounded boxes represent \emph{properties} of these concepts (i.e., an attribute or a characteristic of a concept).

\begin{figure}[ht]
  \centering
  \includegraphics[scale=0.35]{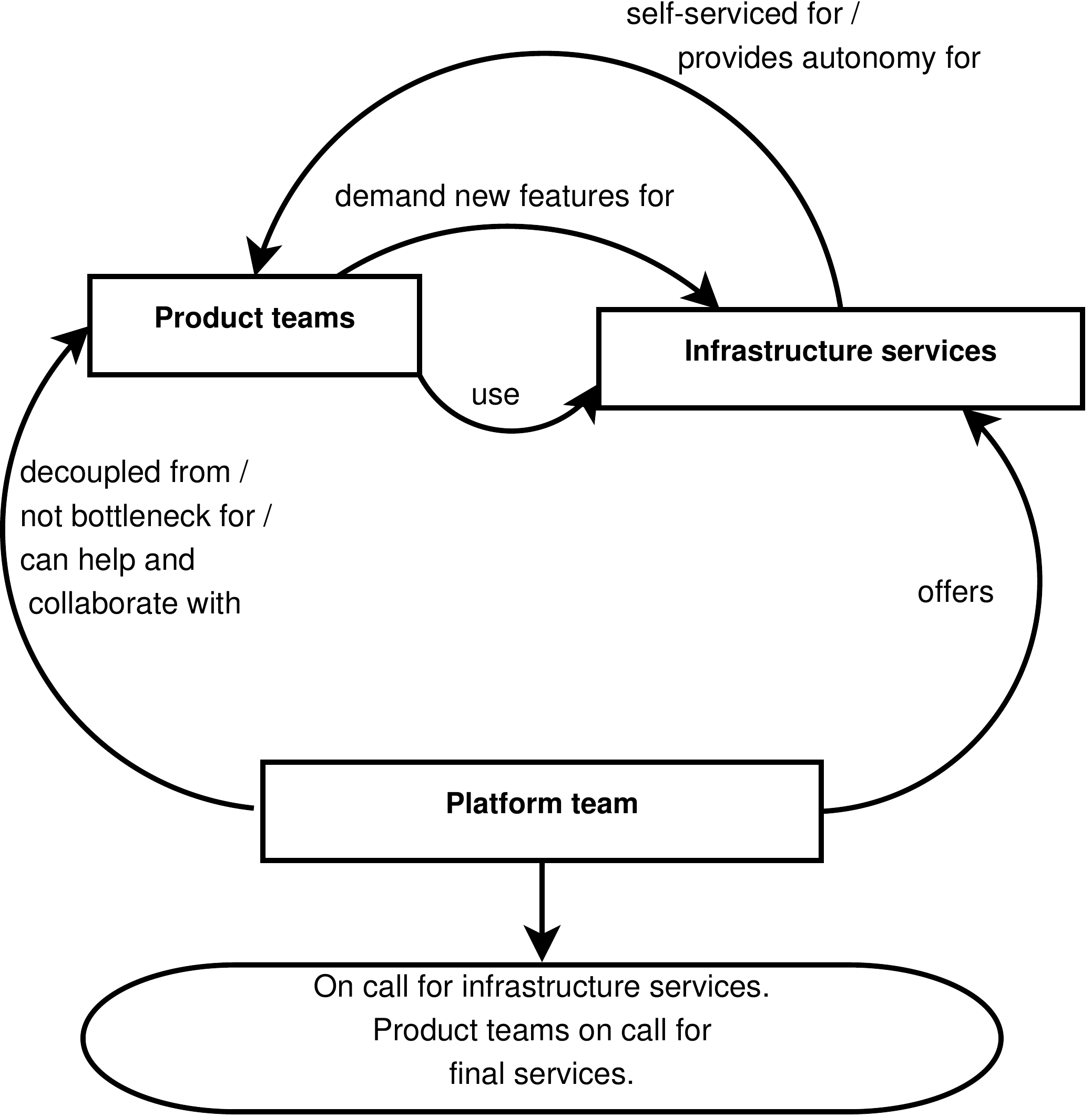}
  \caption{A fragment of our conceptual framework}
  \label{fig:conceptual-framework-fragment}
\end{figure}

As we evolved the conceptual framework and filled our comparison sheet, we developed our theory by classifying each interview by its organizational structure and its supplementary properties. The classification process was based on the concepts provided by the framework and on the analysis of similarities and differences between the interviews, summarized in the comparison sheet. As we evolved our understanding with new interviews, we revisited the classification of previous interviews to refine our theory. For example, after the emergence of a supplementary property, we checked whether we could classify previous interviews with the new property.

After some interviews, one author developed the first version of the coding lists, the comparison sheet, the conceptual framework, and the taxonomy. Based on the transcriptions, other two authors thoroughly reviewed these artifacts, triggering discussions that affected their evolution. When making a decision that could impact our taxonomy, we involved a fourth author. One example of how we evolved and enhanced our taxonomy is how we developed the supplementary properties after twenty interviews. Additional rounds of analysis, discussions, and theory elaboration were undertaken until the submission of this article. We went through this internal review process to reduce the bias of a single researcher performing analysis with preconceived ideas.

\subsection{Theoretical saturation}
\label{sec:saturation}

\begin{figure*}[ht]
  \centering
  \includegraphics[scale=0.6]{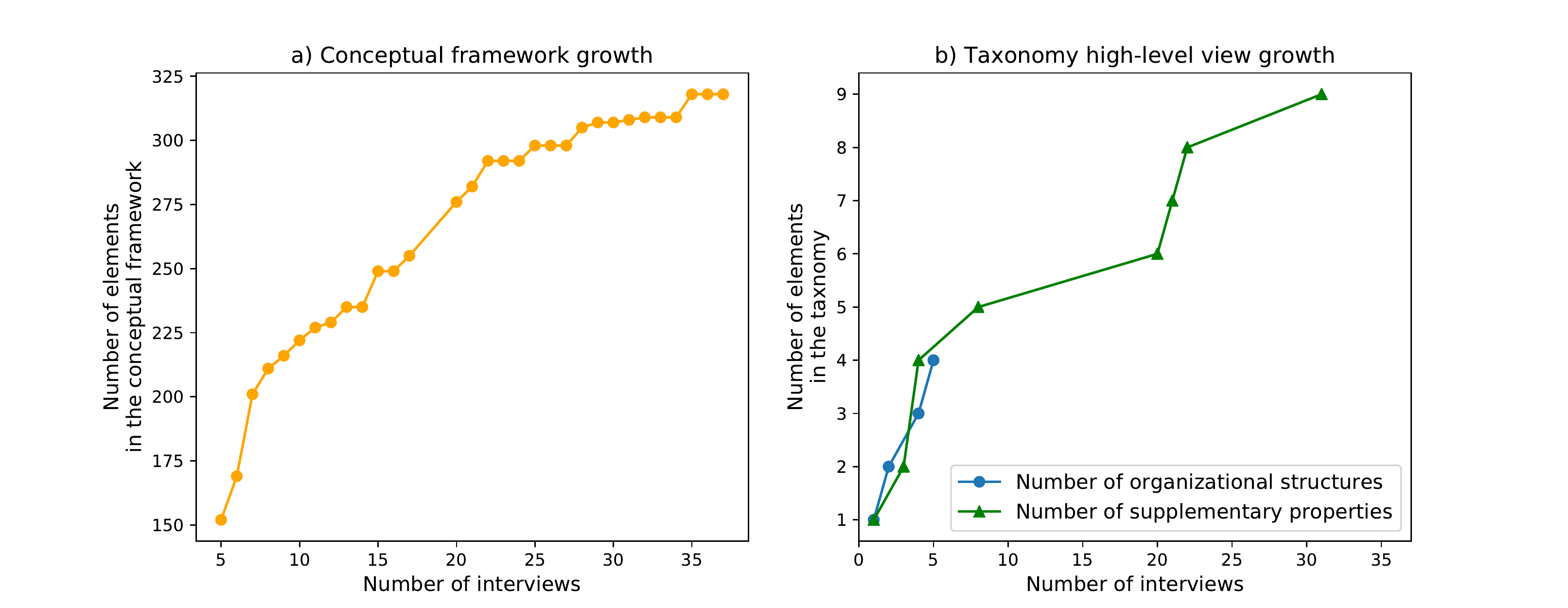}
  \caption{Evidence of theoretical saturation}
  \label{fig:saturation}
\end{figure*}

We consider the size of our conceptual framework as a proxy for how much we have learned so far about our research topic. We define the size of the conceptual framework as the number of elements in the diagram representing it, which counts the number of concepts, conceptual properties, and links. Consider Figure~\ref{fig:saturation}-(a): the $x$ axis represents the number of interviews, while the $y$ axis represents the number of elements in our conceptual framework. In this figure, we see that the last 15 interviews (40\% of them) increased the size of our conceptual framework by only 9\%. Moreover, the last three interviews did not increase the size of the conceptual framework to any degree. This suggests more interviews would provide only a negligible gain for our framework.

In addition to measuring the size of the conceptual framework, an intermediary artifact, we also measured the growth of the taxonomy itself, our final product. Figure~\ref{fig:saturation}-(b) shows that in the first five interviews we discovered our four organizational structures. It also shows that by interview 22, we already had all but one of the supplementary properties. The last discovered supplementary property is an exceptional case, and it was applied only for one interview. Therefore, the decreasing growth of the taxonomy suggests that the last interviews contributed much less to shaping our theory.

By conjoining these two observations of Figure~\ref{fig:saturation}, we claim to have reached enough of saturation for the purposes of our theory.

\subsection{Feedback}
\label{sec:methodology-feedback}

GT aims to formulate a theory that has relevance for practitioners, so it is crucial to also investigate whether findings make sense to them~\cite{ralph2019taxonomies}. Moreover, practitioners can help to identify taxonomy errors, such as inclusion and exclusion errors~\cite{ralph2019taxonomies}. Therefore, we collected feedback on our taxonomy from the study participants, using an online survey\swurl{http://ccsl.ime.usp.br/devops/2020-06-14/feedback-form.html}.
The received feedback is presented in Section~\ref{sec:feedback}.

\section{The Taxonomy of Organizational Structures}
\label{sec:structures}

In this section, we answer our research questions by presenting our taxonomy of the organization of development and infrastructure teams, including the organizational structures we identified, alongside their core and supplementary properties. Core properties are expected to be found in corporations with a given structure. When applicable, we use core properties to discuss delivery performance. Supplementary properties refine the explanation of a structure, but their association with organizations is noncompulsory. 

For each interview, we classified the organizational structure observed. As the differentiation and integration patterns among development and infrastructure may vary for each deployable unit~\cite{shahin2019unit}, it is not possible to assign a single structure to an organization. So the classification is applied according to the interviewee's context. Moreover, we observed in some cases a process of gradually adopting new structural patterns while abdicating from old ones; we classified this as transitions from one organizational structure to another.

\begin{figure*}[ht]
  \centering
  \includegraphics[scale=0.45]{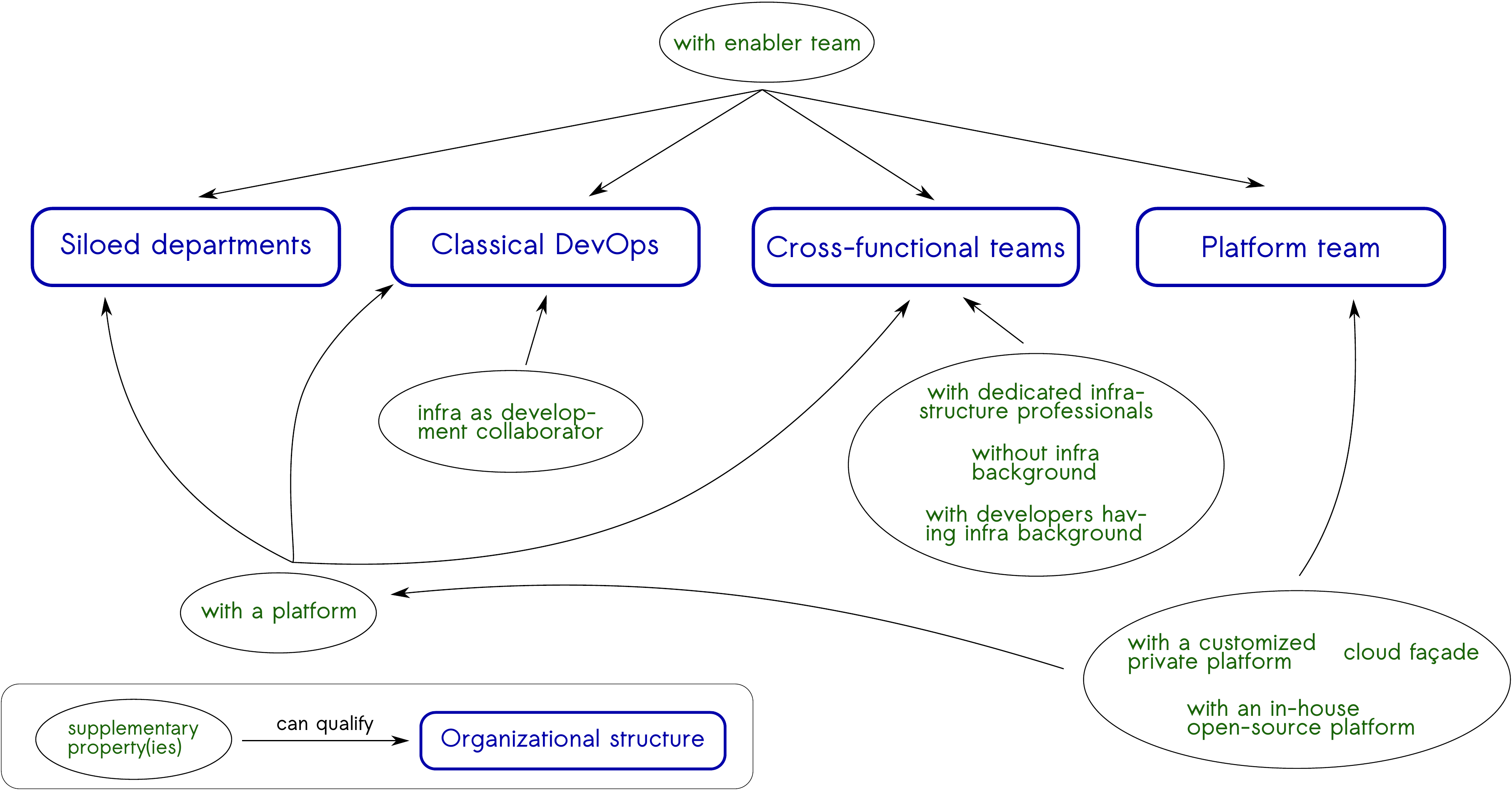}
  \caption{High-level view of our taxonomy: discovered organizational structures and their supplementary properties.}
  \label{fig:taxonomy}
\end{figure*}

Figure~\ref{fig:taxonomy} presents the discovered organizational structures, the primary elements of our taxonomy, alongside the supplementary properties, which qualify the elements pointed out by the arrows. The circles group supplementary properties that can be equally applied for a given element.
Table~\ref{tab:interviews} shows the classification (organizational structure and supplementary properties) applied for each interview, alongside the achieved delivery performance, and the number of employees in the corresponding organization. 

\ifthenelse{\equal{\showtables}{1}}{
\begin{table*}
\footnotesize
  \center
  \caption{Interview's classification. In the second column, ``\textit{to}'' indicates transitioning from one structure to another.}
  \label{tab:interviews}
  \begin{tabular}{ccccc}
    \toprule
    \itshape{Interview} & \itshape{Organizational} & \itshape{Supplementary} & \itshape{Delivery} & \itshape{Organization} \\
     & \itshape{structure} & \itshape{properties} & \itshape{performance} & \itshape{size} \\
    \midrule
    \#I1 & Cross-functional & with dedicated infra professionals & High & $>$ 200 and $<$ 1,000  \\
    \hline
    \#I2 & Classical DevOps &  & High & $>$ 200 and $<$ 1,000  \\
    \hline
    \#I3 & Cross-functional & without infra background & Not-high & $<$ 200 \\
    \hline
    \#I4 & Platform team & cloud fa{\c c}ade & High & $>$ 1,000 \\
         &  & with enabler team &  &  \\
    \hline
    \#I5 & Siloed departments &  & Not-high & $>$ 1,000 \\
    \hline
    \#I6 & Classical DevOps	 &  & Not-high & $>$ 200 and $<$ 1,000  \\
    \hline
    \#I7 & Siloed departments &  & Not-high & $>$ 200 and $<$ 1,000  \\
         & \textit{to} Classical DevOps &  &  &  \\
    \hline
    \#I8 & Siloed departments & with a customized private platform & Not-high & $>$ 1,000 \\
         & \textit{to} Platform team &  &  &  \\
    \hline
    \#I9 & Platform team & cloud fa{\c c}ade & High & $>$ 200 and $<$ 1,000  \\
          &  & with enabler team &  &  \\
    \hline
    \#I10 & Siloed departments &  & Not-high & $<$ 200 \\
    \hline
    \#I11 & Classical DevOps & with enabler team & Not-high & $>$ 200 and $<$ 1,000  \\
    \hline
    \#I12 & Platform team & with a customized private platform & High & $>$ 200 and $<$ 1,000  \\
          &  & with enabler team &  &  \\
    \hline
    \#I13 & Siloed departments &  & Not-high & $>$ 1,000 \\
    \hline
    \#I14 & Classical DevOps & cloud fa{\c c}ade & Not-high & $>$ 200 and $<$ 1,000  \\
          & \textit{to} Platform team &  &  &  \\
    \hline
    \#I15 & Siloed departments &  & Not-high & $>$ 200 and $<$ 1,000  \\
          & \textit{to} Classical DevOps &  &  &  \\
    \hline
    \#I16 & Cross-functional & with dedicated infra professionals & Not-high & $>$ 1,000 \\
          & \textit{to} platform team & with a customized private platform &  &  \\
          & & with enabler team &  &  \\
    \hline
    \#I17 & Classical DevOps & with enabler team & High & $>$ 200 and $<$ 1,000  \\
    \hline
    \#I18 & Classical DevOps & with enabler team & Not-high & $>$ 1,000 \\
    \hline
    \#I19 & Siloed departments &  & Not-high & $<$ 200 \\
    \hline
    \#I20 & Siloed departments & with an in-house open source platform & High & $>$ 1,000 \\
          & \textit{to} Platform team &  &  &  \\
    \hline
    \#I21 & Classical DevOps & with developers having infra background & High & $<$ 200 \\
          & \textit{to} Cross-functional &  &  &  \\
    \hline
    \#I22 & Classical DevOps & with a platform & Not-high & $>$ 1,000 \\
          &  & with a customized private platform &  &  \\
    \hline
    \#I23 & Siloed departments &  & Not-high & $<$ 200 \\
    \hline
    \#I24 & Siloed departments & with dedicated infra professionals & High & $>$ 200 and $<$ 1,000  \\
          & \textit{to} cross-functional &  &  &  \\
    \hline
    \#I25 & Cross-functional & with developers having infra background & Not-high & $<$ 200 \\
          &  & with a platform &  &  \\
          &  & cloud fa{\c c}ade &  &  \\
    \hline
    \#I26 & Siloed departments &  & Not-high & $>$ 1,000 \\
          & \textit{to} Classical DevOps &  &  &  \\
    \hline
    \#I27 & Cross-functional & with dedicated infra professionals & Not-high & $<$ 200 \\
    \hline
    \#I28 & Cross-functional & with dedicated infra professionals & Not-high & $>$ 200 and $<$ 1,000  \\
    \hline
    \#I29 & Classical DevOps &  & High & $<$ 200 \\
    \hline
    \#I30 & Siloed departments & with enabler team & Not-high & $>$ 1,000 \\
          &  & with a platform &  &  \\
          &  & with an in-house open source platform &  &  \\
    \hline
    \#I31 & Classical DevOps & infra as development collaborator & Not-high & $>$ 1,000 \\
          &  & with enabler team &  &  \\
    \hline
    \#I32 & Cross-funcional & without infra background & Not-high & $<$ 200 \\
    \hline
    \#I33 & Platform team & cloud fa{\c c}ade & High & $>$ 1,000 \\
    \hline
    \#I34 & Classical DevOps &  & Not-high & $<$ 200 \\
    \hline
    \#I35 & Cross-functional & with dedicated infra professionals & Not-high & $<$ 200 \\
    \hline
    \#I36 & Classical DevOps &  & Not-high & $>$ 1,000 \\
    \hline
    \#I37 & Siloed departments &  & Not-high & $>$ 1,000 \\
    \bottomrule
  \end{tabular}
\end{table*}
}{TABLE}

Our comprehensive chain of evidence, added as supplementary material, indicates how much of our findings are backed by data we collected. The chain of evidence links each organizational structure and supplementary property to supporting coding, memos, and excerpts. Such linkage is critical for the credibility of qualitative research findings~\cite{yin2010estudodecaso,ralph2019taxonomies,guba1981criteria}.

We now present each one of the organizational structures and their core and supplementary properties.

\subsection{Siloed departments}
\label{sec:siloed}

With siloed departments, developers and the infrastructure staff are segregated from each other, with little direct communication among them. Frictions occurs among silos, since developers want to deliver as much as possible, whereas operations target stability and block deliveries. The DevOps movement was born in 2008~\cite{debois2008devops} to handle such problems. We found seven organizations adhering to this structure, and six others transitioning out of this structure.

While supplementary properties did not emerge for this structure, we found seven core properties for corporations with siloed departments: 

\topic Developers and operators have \textbf{well-defined and differentiated roles}; as stated by \#I20: \excerpt{the wall was very clear: after committing, our work [as developers] was done.} Therefore, there are no conflicts concerning attributions. Well-defined roles and pipelines can decrease the need for inter-departmental direct collaboration (\#I10).

\topic \textbf{Each department is guided by its own interests}, looking for local optimization rather than global optimization, an old and problematic pattern~\cite{goldratt2014goal}. Participant \#I26 told us \excerpt{there is a big war there... the security, governance, and audit groups must still be convinced that the tool [Docker / Kubernetes] is good.}

\topic \textbf{Developers have minimal awareness of what happens in production} (\#I26). So monitoring and handling incidents are mostly done by the infrastructure team (\#I5).

\topic \textbf{Developers often neglect non-functional requirements (NFRs)}, especially security (\#I5). In \#I30, conflicts among developers and the security group arise from disagreement on technical decisions. In other cases, developers have little contact with the security group (\#I26).

\topic \textbf{Limited DevOps initiatives,} centered on adopting tools, do not improve communication and collaboration among teams (\#I30) or spread awareness about automated tests (\#I5, \#I15). In \#I30, a ``DevOps team'' maintaining the deployment pipeline behaves as another silo, sometimes bottlenecking the delivery~\cite{humble2012team}.

\topic \textbf{Organizations are less likely to achieve high delivery performance} as developers need bureaucratic approval to deploy applications and evolve the database schema (\#I5, \#I30). Table~\ref{tab:structures-and-performance} shows that only two of 13 siloed organizations presented high delivery performance, and these two were already transitioning to other structures. However, we observed cases in which low delivery performance was not a problem, such as short-lived research experiments (\#I13) and releases of new phases of a game not requiring code changes (\#I10). Network isolation policies may also hinder frequent deployment (\#B1, \#I7).

\topic We observed a \textbf{lack of proper test automation} in many organizations (\#I5, \#I15, \#I23, \#I26). In \#I26, developers automate only unit tests. Organization \#I15 was leaving test automation only for QA people, which is not suitable for TDD or unit tests. Although siloed organizations are not the only ones that lack test automation (\#I3, \#I32, \#I35), in this structure developers can even ignore its value (\#I5, \#I23, \#I37). We notice that some of the observed scenarios were more challenging for test automation, such as games.

\ifthenelse{\equal{\showtables}{1}}{
\begin{table}[ht]
  \center
  \caption{Organizational structures and delivery performance observed in our interviews.}
  \label{tab:structures-and-performance}
  \begin{tabular}{lll}
    \toprule
    \itshape{Organizational} & \itshape{Delivery} & \itshape{Number of} \\
    \itshape{structure} & \itshape{performance} & \itshape{interviews} \\
    \midrule
    Siloed departments & Not-high & 7\\
    \hline
    Classical DevOps & High & 3\\
    \hline
    Classical DevOps & Not-high & 7\\
    \hline
    Cross-functional & High & 1\\
    \hline
    Cross-functional & Not-high & 6\\
    \hline
    Platform team & High & 4\\
    \hline
    Siloed departments & Not-high & 3\\
    \textit{to} Classical DevOps & & \\
    \hline
    Siloed departments & High & 1\\
    \textit{to} Cross-functional & & \\
    \hline
    Siloed departments & High & 1\\
    \textit{to} Platform team & & \\
    \hline
    Siloed departments & Not-high & 1\\
    \textit{to} Platform team & & \\
    \hline
    Classical DevOps & High & 1\\
    \textit{to} Cross-functional & & \\
    \hline
    Classical DevOps & Not-high & 1\\
     \textit{to} Platform team & & \\
    \hline
    Cross-functional & Not-high & 1\\
    \textit{to} Platform team & & \\
  \bottomrule
\end{tabular}
\end{table}
}{TABLE}

\subsection{Classical DevOps}
\label{sec:classical-devops}

The classical DevOps structure focuses on collaboration among developers and the infrastructure team. It does not eliminate all conflicts, but promotes a better environment to deal with them (\#I34). We named this structure ``Classical DevOps'' because we understand that a collaborative culture is the core DevOps concern~\cite{luz2019devops,leite2019survey,davis2016effective}. We classified ten organizations into this structure. We also observed three organizations transitioning to this structure and three transitioning out of this structure.

The eight core properties observed for organizations adopting classical DevOps are as follows:

\topic We observed that, in classical DevOps settings, many practices foster \textbf{a culture of collaboration}. We saw the sharing of database management: infrastructure staff creates and fine tunes the database, whereas developers write queries and manage the database schema (\#I17). We heard about open communication among developers and the infrastructure team (\#I2, \#I6, \#I17, \#I22, \#I31, \#I36). Participant \#I2 highlighted that: \excerpt{Development and infrastructure teams participate in the same chat; it even looks like everyone is part of the same team.} Developers also support the product in its initial production (\#I31). 

\topic Roles remain well-defined, and despite the collaboration on some activities, there are usually \textbf{no conflicts over who is responsible for each task}.

\topic Developers feel relieved when they can rely on the infrastructure team (\#I17). Participant \#I31 claimed that his previous job in a cross-functional team had a much more stressful environment than his current position in a development team in a classical DevOps environment. On the other hand, \textbf{stress can persist at high levels for the infrastructure team} (\#I34), especially \excerpt{if the application is ill-designed and has low performance} (\#I36).

\topic In this structure, the project's success depends on the \textbf{alignment of different departments}, which is not trivial to achieve. In \#B3, different teams understood the organization's goals and the consequences of not solving problems, like wrongly computing amounts in the order of millions of dollars. Moreover, \#I7 described that alignment emerges when employees focus on problem-solving rather than role attributions.

\topic \textbf{Development and infrastructure teams share NFR responsibilities} (\#17). For example, in \#I2, both were very concerned with low latency, a primary requirement for their application.

\topic Usually, \textbf{the infrastructure staff is the front line of tracking monitoring and incident handling} (\#I2, \#I11, \#I29, \#I31, \#I36). However, if needed, developers are summoned and collaborate (\#I17, \#I34). In \#I34, monitoring alerts are directed to the infrastructure team but copied to developers. However, in some cases developers never work after-hours (\#I2, \#I22).


\topic Humble expects a culture of collaboration among developers and the infrastructure staff to \textbf{prescind from a ``DevOps team''}~\cite{humble2012team}. We understand this criticism applies to DevOps teams with dedicated members, such as we saw in \#I30, since they behave as new silos. However, we found in \#I36 a well-running DevOps team working as a committee for strategic decisions --- a forum for the leadership of different departments. We also found DevOps groups working as guilds (\#I4, \#I8), supporting knowledge exchange among different departments~\cite{spotify2014culture}.

\topic Collaboration and delivery automation, critical values of the DevOps movement, are \textbf{not enough to achieve high delivery performance}. Of 10 classical DevOps organizations not transitioning from or to other structures, only three presented high delivery performance (Table~\ref{tab:structures-and-performance}). One possible reason is the lack of proper test automation (\#I22, \#I36)\cite{shahin2020architecture}. Another limitation for delivery performance is the adoption of release windows (\#I11, \#I31, \#I14, \#I36), which seek to mitigate deployment risk by restricting software delivery to periodic time slots. Release windows are adopted by considering either the massive number of users (\#I31) or the system's financial criticality (\#I36). Release windows may also result from fragile architectures (\#I37) or the monolith architectural style (\#I11) since any deployment has an increased risk of affecting the whole system. 

\vspace{0.2cm}
\noindent
\emph{Supplementary properties}

For classical DevOps organizations, we found one supplementary property that we describe in the following.

\vspace{0.1cm}
\textbf{Infra as development collaborator.} The infrastructure staff contributes to the application code to optimize the system's performance, reliability, stability, and availability. Although this aptitude requires advanced coding skills from infrastructure professionals, it is a suitable strategy for maintaining large-scale systems, like the ones owned by \#I31.

\subsection{Cross-functional teams}
\label{sec:cross-functional}

In our context, a cross-functional team takes responsibility for both software development and infrastructure management. This structure aligns with the Amazon motto ``\emph{You built it, you run it}''~\cite{gray2006conversation} and with the ``autonomous squads'' at Spotify~\cite{spotify2014culture}. This gives more freedom to the team, along with a great deal of responsibility. As interviewee \#I1 described: \excerpt{it's like each team is a different mini-company, having the freedom to manage its own budget and infrastructure.} We found seven organizations with this structure, two organizations transitioning to this structure, and one transitioning out of it.

The four core properties found for cross-functional teams are as follows:

\topic \textbf{Independence among teams may lead to misalignment.} Lack of communication and standardization among cross-functional teams within a single organization may lead to duplicated efforts (\#I28). However, this is not always a problem (\#I1). 

\topic \textbf{It is hard to ensure a team has all the necessary skills.} For instance, we interviewed two cross-functional teams with no infrastructure expertise (\#I3, \#I32). Participant \#I27 recognizes that \excerpt{there is a lack of knowledge} on infrastructure, deployment automation, and monitoring. A possible reason for such adversity is that, as \#I29 taught us, it is hard to hire infrastructure specialists and senior developers.

\topic We expected cross-functional teams to provide too much idle time for specialists, as opposed to centralized pools of specialization. However, we find \textbf{no evidence of idleness for specialists}. From \#I16, we heard quite the opposite: the infrastructure specialists were too busy to be shared with other teams. Having the infrastructure specialists code features in their spare time avoids such idleness (\#I35).

\topic \textbf{Most of the cross-functional teams we interviewed were in small organizations} (Table~\ref{tab:structures-and-sizes}), likely because there is no sense in creating multiple teams in too small organizations.

\ifthenelse{\equal{\showtables}{1}}{
\begin{table}[ht]
  \center
  \caption{Organizational structures and organization size observed in our interviews.}
  \label{tab:structures-and-sizes}
  \begin{tabular}{lll}
    \toprule
    \itshape{Organizational} & \itshape{Organization} & \itshape{Number of} \\
    \itshape{structure} & \itshape{size} & \itshape{interviews} \\
    \midrule
    Siloed departments & $<$ 200 & 3\\
    \hline
    Siloed departments & $>$ 1,000 & 4\\
    \hline
    Classical DevOps & $<$ 200 & 2\\
    \hline
    Classical DevOps & $>$ 200 and $<$ 1,000  & 4\\
    \hline
    Classical DevOps & $>$ 1,000 & 4\\
    \hline
    Cross-functional & $<$ 200 & 5\\
    \hline
    Cross-functional & $>$ 200 and $<$ 1,000  & 2\\
    \hline
    Platform team & $>$ 200 and $<$ 1,000  & 2\\
    \hline
    Platform team & $>$ 1,000 & 2\\
    \hline
    Siloed departments & $>$ 200 and $<$ 1,000  & 2\\
    \textit{to} Classical DevOps & & \\
    \hline
    Siloed departments & $>$ 1,000 & 1\\
    \textit{to} Classical DevOps & & \\
    \hline
    Siloed departments & $>$ 200 and $<$ 1,000  & 1\\
    \textit{to} Cross-functional & & \\
    \hline
    Siloed departments & $>$ 1,000 & 2\\
    \textit{to} Platform team & & \\
    \hline
    Classical DevOps & $<$ 200 & 1\\
    \textit{to} Cross-functional & & \\
    \hline
    Classical DevOps & $>$ 200 and $<$ 1,000  & 1\\
    \textit{to} Platform team & & \\
    \hline
    Cross-functional & $>$ 1,000 & 1\\
    \textit{to} Platform team & & \\
  \bottomrule
\end{tabular}
\end{table}
}{TABLE}

\vspace{0.2cm}
\noindent
\emph{Supplementary properties}

\vspace{0.1cm}
\textbf{Dedicated infra professionals.} The team has specialized people dedicated to infrastructure tasks. In \#I1, one employee specializes in physical infrastructure, and another is ``the DevOps'', taking care of the deployment pipeline and monitoring. In this circumstance, the infrastructure specialists become the front-line for tackling incidents and monitoring (\#I28, \#I35).

\vspace{0.1cm}
\textbf{Developers with infra background.} The team has developers knowledgeable in infrastructure management; these professionals are also called full-stack engineers or even DevOps engineers (\#I25). Participant \#I25 is a full-stack engineer and claimed to \excerpt{know all the involved technologies: front-end, back-end, and infrastructure; so I'm the person able to link all of them and to firefight when needed.}
Participant \#I29, a consultant, is skeptical regarding full-stack engineers and stated that \excerpt{these people are not up to the task.} He complained that developers are usually unaware of how to fine tune the application, such as configuring database connections.

\vspace{0.1cm}
\textbf{No infra background.} Product teams manage the infrastructure without the corresponding expertise. We saw this pattern in two places. One was a very small company and had just released their application, having only a few users (\#I32) and being uncertain about hiring specialized people soon. Interviewee \#I3 understands that operations work (e.g., spotting errors during firmware updates in IoT devices and starting Amazon VMs for new clients) is too menial for software engineers, taking too much of their expensive time. So the organization was planning the creation of an operations sector composed of a cheaper workforce. Interviewee \#I19 argued that such an arrangement could not sustain growth in his company in the past.

\subsection{Platform teams}
\label{sec:platform}

Platform teams are infrastructure teams that provide highly automated infrastructure services that can be self-serviced by developers for application deployment. The infrastructure team is no longer a ``support team''; it behaves like a product team, with the ``platform'' as its product and developers as internal customers. In this setting, infrastructure specialists need coding skills; product teams have to operate their business services; and the platform handles much of the non-functional concerns. We found four organizations fully embracing this model and four others in the process of adopting it. We also observed the platform team pattern in three of the brainstorming sessions.

The core properties of platform teams are as follows:

\topic \textbf{Product teams are fully accountable for the non-functional requirements of their services.} They become the first ones called when there is an incident, which is escalated to the infrastructure people only if the problem relates to an infrastructure service (\#I8, \#I9, \#I12, \#I33). 

\topic Although the product team becomes fully responsible for NFRs of its services, it is not a significant burden that developers try to refuse (\#I33). \textbf{The platform itself handles many NFR concerns}, such as load balancing, auto-scaling, throttling, and high-speed communications between data-centers (\#I4, \#I8, \#I16, \#I33). As participant \#I33 told us, \excerpt{you don't need to worry about how things work, they just work.} Moreover, we observed infrastructure people willingly supporting developers for the sake of services availability, performance, and security (\#I9, \#I14).

\topic \textbf{Product teams become decoupled from the members of the platform team.} Usually, the communication among these teams happens when developers and infrastructure people gather to solve incidents (\#I8, \#I9); when infrastructure people provide consulting for developers to master non-functional concerns (\#I9); or when developers demand new capabilities from the platform (\#I8, \#I12). In this way, the decoupling between the platform and product teams does not imply the absence of collaboration among these groups.

\topic \textbf{The infrastructure team is no longer requested for operational tasks.} The operational tasks are automated by the platform. Therefore, one cannot merely call platform-team members ``operators,'' since they also engineer the infrastructure solution. We remark that, in other industries, ``operator'' is a title attributed to menial workers.

\topic The platform \textbf{avoids the need for product teams to have infrastructure specialists}.
Participant \#I33 expressed wanting to better understand what happens \excerpt{under the hood} of the platform, which indicates how well the platform relieves the team from mastering infrastructure concerns. On the other hand, since developers are responsible for the deployment, they must have some basic knowledge about the infrastructure and the platform itself.

\topic \textbf{The platform may not be enough to deal with particular requirements.} Participant \#I16 stated that \excerpt{if a lot of people do similar functionality, over time usually it gets integrated to the platform... but each team will have something very specialized...} to explain the presence of infrastructure staff within the team, even with the usage of a platform, considering the massive number of virtual machines to be managed.

\topic If the organization develops a new platform to deal with its specificities, \textbf{it will require development skills from the infrastructure team}. Nevertheless, even without developing a new platform, the infrastructure team must have a ``dev mindset'' to produce scripts and use infrastructure-as-code~\cite{morris2016infraascode} to automate the delivery path (\#I14). One strategy we observed to meet this need was to hire previous developers for the infrastructure team (\#I14).

\topic All four organizations that have fully embraced the \textbf{platform team structure are high performers}, while no other structure provided such a level of success (Table~\ref{tab:structures-and-performance}). An explanation for such a relation is that this structure decouples the infrastructure and product teams, which prevents the infrastructure team from bottlenecking the delivery path. As stated by \#I20: \excerpt{Now developers have autonomy for going from zero to production without having to wait for anyone.} This structure also contributes to service reliability by letting product teams handle non-functional requirements and incidents. 

\topic In Table~\ref{tab:structures-and-sizes}, among the interviewed organizations with platform teams, none had less than 200 employees. Since assembling a platform team requires dedicated staff with specialized knowledge, it makes sense that such a structure is \textbf{not suitable for small companies}.

\vspace{0.2cm}
\noindent
\emph{Supplementary properties}

The description of a platform can be refined by applying one of the following supplementary properties:

\vspace{0.1cm}
\textbf{Cloud fa\c{c}ade.} The platform ultimately deploys applications on public clouds, such as Amazon WS, Google Cloud, or Azure. Although these clouds allow easier deployment when compared to managing physical servers, they still offer dozens of services and a multitude of configurations. The in-house platform standardizes the usage of public cloud vendors within the organization, so developers do not need to understand many details about the cloud (\#I4, \#I14, \#I33); therefore \excerpt{enhancing the usability of the [cloud] infrastructure} (\#I16).


\vspace{0.1cm}
\textbf{Customized private platform.} The platform is built on top of internal physical servers (\#B1, \#I1, \#I8, \#I12, \#I20), hiding infrastructure complexities from developers, such as the use and even the existence of Kubernetes, an open-source platform used for managing the lifecycle of Docker containers.

\vspace{0.1cm}
\textbf{In-house open-source platform.} The platform is an open-source software deployed on-premise. Organization \#I20 uses Rancher\swurl{http://rancher.com}, a graphical interface for developers to interact with Kubernetes.

\subsection{Shared supplementary properties}

This section presents the found supplementary properties that are not linked to one organizational structure only. These properties are relevant since they are shared among multiple organization structures, as depicted in Figure~\ref{fig:taxonomy}.

\vspace{0.1cm}
\textbf{Enabler team.} An enabler team provides consulting and tools for product teams but does not own any service. Consulting can be on performance (\#I18) or security (\#I9, \#I16, \#I31), for example. Tools provided by enabler teams include the deployment pipeline (\#I4, \#I30), high-availability mechanisms (\#I11), monitoring tools (\#I12), and security tools (\#I17). We found them in every organizational structure. We learned the term ``enabler team'' during our interview with \#I11.

\ifthenelse{\equal{\showtables}{1}}{
\begin{table*}
\footnotesize
  \center
  \caption{Summary of organizational structures.}
  \label{tab:structures}
  \begin{tabular}{l|l l l}
    \toprule
    \itshape{Organizational structure} & \itshape{Development differentiation} & \itshape{Infrastructure differentiation} & \itshape{Integration} \\
    \hline
    Siloed departments & Just builds the & Responsible for all & Limited collaboration   \\
                       & application package & operations activities & among the groups \\
    \hline
    Classical DevOps & Participates/collaborates in & Responsible for all & Intense collaboration \\
                     & some operations activities & operations activities & among the groups \\
    \hline
    Cross-functional teams & Responsible for all & Does not exist & ---  \\
                           & operations activities & & \\
    \hline
    Platform teams & Responsible for all & Provides the platform & Interaction happens \\
                   & operations activities & automating much of & for specific situations, \\
                   & with the platform support & the operations activities & not on a daily basis \\
    \bottomrule
  \end{tabular}
\end{table*}
}{TABLE}

\vspace{0.1cm}
\textbf{With a platform.} The organization possesses a platform that can provide deployment automation, but not following the patterns of human interaction and collaboration described by the core properties of platform teams. Participant \#I25 developed an \excerpt{autonomous IaaC for integration and deployment with Google Cloud,} which provides a platform's capabilities to other developers of the team. However, since in this context there is a single cross-functional team, it cannot be called a ``platform team.'' We classified organization \#I30 as a siloed structure, even with a platform team, since developers and the platform team have a conflicted relationship. The supplementary properties of platform teams can also be applied to organizations with a platform.

\subsection{Transitioning}

\begin{figure}[ht]
  \centering
  \includegraphics[scale=0.4]{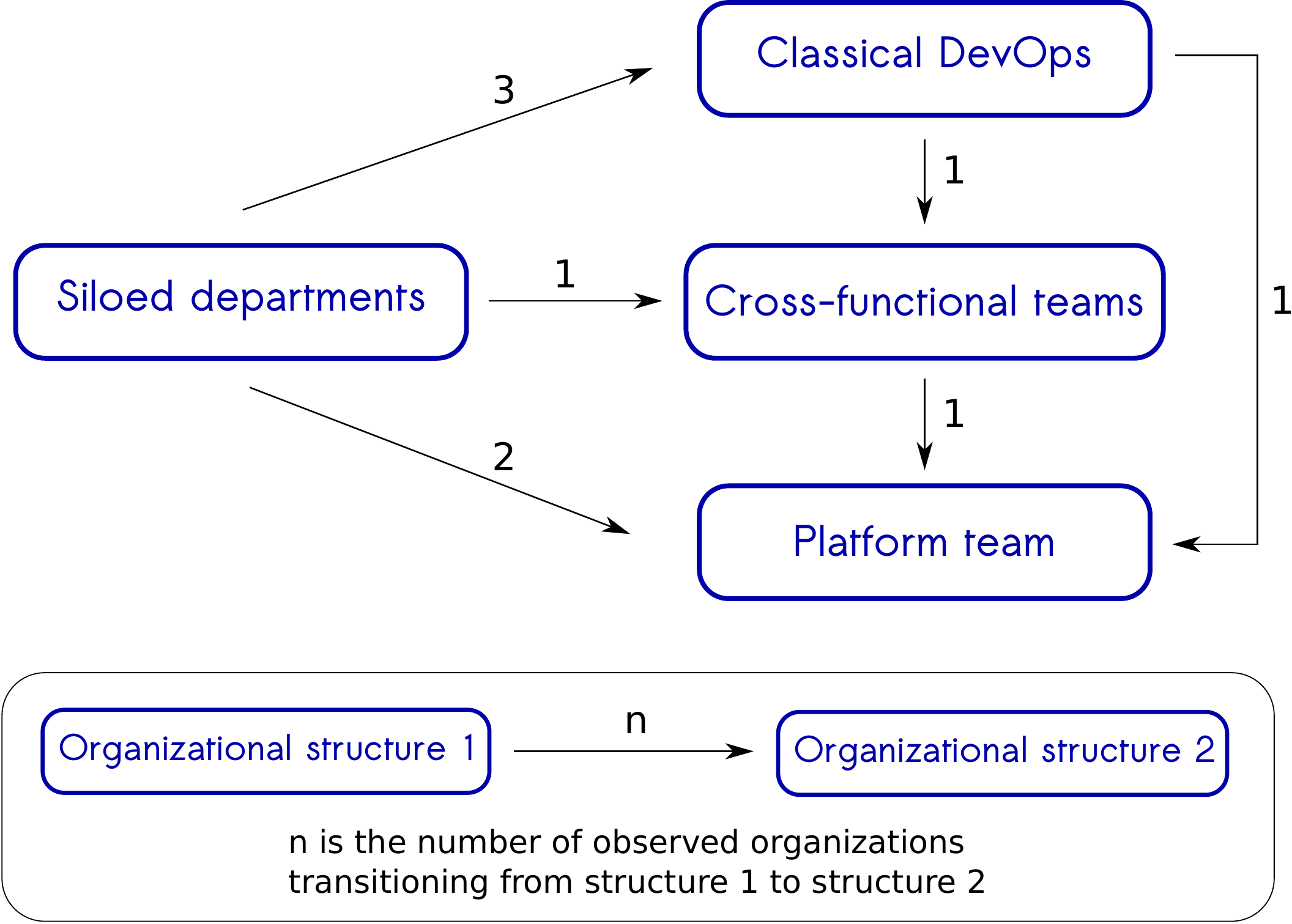}
  \caption{Observed transition flows}
  \label{fig:transitions}
\end{figure}

Organizations are not static. We identified nine of them transitioning from one structure to another. Considering the transition flows in Figure~\ref{fig:transitions}, we perceive that \emph{i)} no organization is transitioning \emph{to} siloed departments, \emph{ii)} most of the transitions are \emph{from} siloed departments, and \emph{iii)} no organization is transitioning \emph{out of the} platform team. These observations agree with our theoretical considerations about the problems of siloed structures and the promises of platform teams.

Nonetheless, transitioning organizational structures is a hard endeavor, as confirmed by some interviewees. Although his organization did an excellent job transitioning to a platform structure and achieving high-delivery performance, interviewee \#I20 claims that the \excerpt{old world} still coexists with the \excerpt{new one}. In the same way, as reported by Nybom \emph{et al.}~\cite{nybom2016mixing}, there are some responsibility conflicts and \excerpt{dissident forces}: some operations personnel do not like developers with administrative powers, while some developers do not want such powers. The interviewee declared that \excerpt{it's not yet everybody together.}

Similarly, interviewee \#21 stated that \excerpt{There are two worlds... one was born in the cloud, and it's nice that it influences the legacy system to become more robust. There are many worlds we wish to bring together. However, we need to rewrite even the culture; we must reset everything.} These examples also show how culture is a crucial factor for change.

\subsection{Summary}

We close the description of our taxonomy by highlighting the key differences among its organizational structures. Table~\ref{tab:structures} summarizes, for each structure, \emph{(i)} the differentiation between development and infrastructure groups regarding operations activities (deployment, infrastructure setup, and service operation in run-time); and  \emph{(ii)} how these groups interact (integration).

\section{Discussion}
\label{sec:discussion}

In this section, we discuss feedback sessions and our emerging theory in light of the received feedback.

\subsection{Feedback}
\label{sec:feedback}

We sent to each one of the 37 first interviewees a feedback form asking whether the interviewee agreed with the chosen classification (organizational structure and supplementary properties) for its context using the following Likert scale: strongly agree, weakly agree, I do not know, weakly disagree, strongly disagree. In case of disagreement, there were free text fields for explanation. We also asked the interviewees whether they perceived our taxonomy as comprehensive and whether they would add or remove elements. Finally, we also left a free field for general comments. We sent the form in four batches of twenty, five, five, and seven emails spread across the last five months of our interviewing period; we used the feedback incrementally to refine our theory. We also attached to the emails a digest describing our taxonomy.

We received 11 answers. Nine participants strongly agreed with the received classification regarding the organizational structure, while two of them weakly agreed with it. No one disagreed. Five participants strongly agreed with the received classification regarding the supplementary properties, while one of them weakly agreed with it. Regarding our model's comprehensiveness, seven participants strongly agreed with it, three of them weakly agreed with it, and one of them did not know. This result suggests resonance of the participants with our theory, which refers to the
degree to which findings make sense to participants.

The free-text answers from participants were valuable in refining the taxonomy. We conceived the supplementary properties from the analysis of the first round of feedback. One interviewee's comments helped us improve our taxonomy digest (we refined a figure to better express the idea of platform team). Moreover, some participants raised concerns about how different parts of the organization act under different patterns, and how this evolves. We considered such concerns since our classification is not for the whole organization, but for the interviewee's context at a point in time.

\subsection{Evaluating our Emerging Theory}
\label{sec:evaluation}

According to Ralph, good taxonomies should increase cognitive efficiency and assist reasoning by facilitating more inferences~\cite{ralph2019taxonomies}. However, evaluating whether a taxonomy satisfies such quality criteria demands a different type of research approach than we employed; e.g., a case study~\cite{yin2010estudodecaso}. Although case studies can test theories~\cite{eisenhardt1989theory}, they are more useful for demonstrating how a current theory is incomplete or inadequate to explain the observed case~\cite{pinfield1986decision,anderson1983cuban}. A single case study does not prove a social theory. Moreover, even after decades of robust tests confirming theories, new work can still expand them~\cite{sirmon2011resource}.

On the other hand, grounded Theory (GT) focuses on generating theory rather than validating preconceived hypotheses. Although researchers can be tempted to try to validate their theory as soon as it is born, Glaser and Strauss caution that verificational approaches hinder theory development too early~\cite{glaser1999gt}.
Therefore, in this paper, we present only incipient steps related to theory assessment: hearing feedback on our taxonomy (Section~\ref{sec:feedback}) and comparing it to other existing classifications (Section~\ref{sec:related-work}). Nonetheless, such steps evidence the resonance of our theory.

By following Guba's framework for naturalistic research evaluation~\cite{guba1981criteria}, practitioner and literature resonance also provide some credibility (internal validity, how plausible or true the findings are) and confirmability (objectivity, opportunities for correcting research bias). The remaining of Guba's criteria are dependability (reliability), which is provided by our chain of evidence, and transferability (external validity or generalizability), which is supported by our diverse selection of participants.

Though it is inadequate to evaluate our emerging theory as a finished product, we can \emph{evaluate our process of generating the theory}~\cite{glaser1999gt, ralph2019taxonomies}. This can be done by checking adherence to GT prescriptions and Ralph's recommendations for taxonomy generation~\cite{ralph2019taxonomies}. We described the adherence to GT guidelines (i.e., theoretical sampling, coding, theoretical sensitivity, and theoretical saturation) in Section~\ref{sec:research-design}. In the following, we discuss our adherence to Ralph's recommendations.

A warning from Ralph is that theory should explain, not prescribe. In this way, it is essential to note that although our theory is intended to be used by practitioners in practical settings, as is the goal of a grounded theory, the theory itself provides an explanation of the world, not a guide for action.

A more severe impact on our work comes from Ralph's advice for favoring first-hand observations over interviews due to interviewees' biases. Although we acknowledge his concerns, in our case organizational structures are too abstract to grasp only by observation, even meetings observation, without further conversations with observed people. Our context thus differs from other software engineering situations, such as observing a pair-programming session. Nonetheless, anonymity reduces these biases by favoring an open attitude from the participants. We took care not to ask the research questions directly to participants, which would force our preconceptions onto them~\cite{ralph2019taxonomies}. We carefully crafted second-level questions~\cite{yin2010estudodecaso}, which are more objective than the research questions. Our interview protocol\swurl{http://ccsl.ime.usp.br/devops/2020-06-14/interview-protocol.html} presents each interview question, followed by its rationale.

We acknowledge the importance of triangulating results with other kinds of data and with other participants in the same organizations. However, we spare these strategies for the future phase of our research. Observing more scenarios is more valuable for this initial phase of theory elaboration than interviewing more people in each organization, which would lead to interviews in fewer companies.

Ralph still cautions that researchers should take care in selecting evaluation criteria to assess a theory development. Since multiple criteria exist and there is no universally accepted set of criteria, researchers must choose criteria that make sense for the emerging theory~\cite{ralph2019taxonomies}. Nonetheless, we apply to our research three crucial criteria suggested by Ralph: \emph{(i)} the empirical study is well-executed and clearly described, \emph{(ii)} there is an explicit chain of evidence from results back to supporting data (which does not exclude the existence of non-replicable intuitive leaps~\cite{ralph2019taxonomies}), and \emph{(iii)} the need for the proposed theory must be clear. The present text and the complete chain of evidence, provided as supplementary material, must suffice to the reader to judge these concerns.

\section{Related work}
\label{sec:related-work}

Recent research has discussed the benefits and challenges of continuous delivery~\cite{Chen:IEEESW:2015,leppanen2015highways,olsson2012stairway,Schermann:2016:ICPC,neely2013easy}.
Among the challenges, Chen~\emph{et al.} include organizational issues related to tensions among groups within an organization~\cite{Chen:IEEESW:2015}, which demands studies on the organization of teams. Some of these studies have focused on how developers migrate from teams or companies~\cite{mockus2010volatility,santos2016jobrotation,hilton2018organizational}. 
Nonetheless, the literature about the inter-team arrangements for managing IT infrastructure in a continuous delivery context is still limited. In the following, we discuss this existing literature~\cite{humble2011devops,nybom2016mixing,puppet2018devops,skelton2013topologies,
skelton2019teamtopologies,shahin2017structures} by comparing their classifications of inter-team relations to our taxonomy.

In one of the foundational writings on DevOps~\cite{humble2011devops}, Humble and Molesky start by criticizing the siloed department structure and management by project. They then follow by advocating cross-functional teams and management by product. However, they also suggest practices for strengthening collaboration among development and operations, which makes sense in the classical DevOps structure. Such practices include operators attending agile ceremonies and developers contributing to incident solving. Humble and Molesky also envision operation groups offering support services (e.g., continuous integration) and infrastructure as a service to product teams, which relates in our taxonomy to enabler teams and the platform team structure.

Nybom et al. present three distinct approaches to DevOps adoption~\cite{nybom2016mixing}: (i) assigning development and operations responsibilities to all engineers; (ii) composing cross-functional teams of developers and operators; and (iii) creating a DevOps team to bridge development and operations. However, the article is about a case study matching the first approach only; developers undertook operational tasks, and collaboration was promoted among the development and operations departments. According to our taxonomy, such a scenario was an attempt to migrate from siloed departments to classical DevOps. However, despite some perceived benefits, new sources of friction emerged among the departments, and several employees disagreed with the adopted approach. We associate these sub-optimal results to the reported lack of automation investments, which suggests that trying any DevOps adoption without aiming for continuous delivery is not promising.

The 2018 State of DevOps Report surveyed respondents about the organizational structures used in their DevOps journeys~\cite{puppet2018devops}, offering a closed set of alternatives: \emph{cross-functional teams responsible for specific services or applications}, \emph{dedicated DevOps teams}, \emph{centralized IT teams with multiple application development teams}, \emph{site reliability engineering teams}, and \emph{service teams providing DevOps capabilities (e.g., building test environments, monitoring)}. However, the text does not further describe such options. Thus, associating our structures to the options presented by the survey would be an error-prone activity.

Skelton and Pais present nine ``DevOps topologies'' and seven anti-patterns~\cite{skelton2013topologies}, as the most informal of our comparison sources -- a blog post. The presentation of each topology and anti-pattern is short, lacking details about how corporations apply them. 
We now present the correspondences among the DevOps topologies / anti-patterns and our taxonomy. \emph{Dev and ops silos} corresponds to our siloed departments structure. \emph{Dev don't need ops} corresponds to our cross-functional teams with no infra background. In some organizations adopting classical DevOps, we saw the rebranding of the infrastructure team to SRE or DevOps (\#I2, \#I6, \#I31), but this situation did not entirely match the \emph{rebranded sysadmin} anti-pattern since there were cultural changes in the observed cases. \emph{Ops embedded in dev team} corresponds to our cross-functional teams with dedicated infra professionals, although we saw positive results with this configuration in \#I1. In a siloed organization (\#I30), we also observed the \emph{DevOps team silo}.
\emph{Dev and ops collaboration} corresponds to our classical DevOps structure. \emph{Ops as infrastructure-as-a-service (platform)} corresponds to our platform teams. We consider \emph{SRE team} topology to match our classical DevOps structures with infra as development collaborator, although the topology description does not include this SRE activity of coding the application to improve its non-functional requirements~\cite{beyer2016sre}.

The \emph{Team Topologies} book~\cite{skelton2019teamtopologies}, from the same authors of the DevOps topologies blog post, presents four dynamic patterns of teams for software-producing corporations:\\ \emph{stream-aligned teams}, delivering software in a business-aligned constant flow; \emph{complicated sub-system teams}, with highly specialized people working on a complicated problem; \emph{enabler teams}, providing consulting for stream-aligned teams in a specific technical or product domain; and \emph{platform teams}, providing internal services for self-service by stream-aligned teams, abstracting infrastructure and increasing autonomy for stream-aligned teams.

The \emph{stream-aligned team} corresponds to what we call a product team or development team in this paper. The \emph{complicated sub-system team} is not considered in our taxonomy since it is related to splitting work within development only. The \emph{enabler team} proposed by Skelton and Pais is very close to what we also call an enabler team (e.g., interviewee \#I18 was in an enabler team providing consulting on performance for product teams); however, Skelton and Pais advocate that such consulting must be time-bounded, which is an aspect absent from our observed enabler teams. Finally, the \emph{platform team} proposed by Skelton and Pais includes our notion of platform team; however, our concept is restricted to the services related to the execution environment of the applications, i.e., while we considered teams providing pipeline services as enabler teams, Skelton and Pais would consider them as platform teams.

Another significant difference from team topologies to our work is that the book seeks to present \emph{things how they should be}, while we try to summarize \emph{thing how they are}. In this sense, although we acknowledge the existence of the \emph{classical DevOps} structure, it is not included in the team topologies, since the authors discourage the handover it causes. We also note that the terms ``platform team'' and ``enabler team'' emerged from our interviewees, without the Teams Topologies book's direct influence.

Most of the discussed work so far~\cite{nybom2016mixing,humble2011devops,puppet2018devops,skelton2013topologies} presents sets of organizational structures without an empirical elaboration of how such sets were conceived. The \emph{Team Topologies} book suggests that the proposed topologies emerged from field observations, but lacked a scientific methodology. In this way, Shahin~\emph{et al.}~\cite{shahin2017structures} present the closest work to ours, with a set of structures based on field data and scientific guidelines.

Shahin~\emph{et al.}~\cite{shahin2017structures} conducted semi-structured interviews in 19 organizations and surveyed 93 practitioners to empirically investigate how development and operation teams are organized in the software industry toward adopting continuous delivery practices. They found four types of team structures: \emph{i) separate Dev and Ops teams with higher collaboration}, \emph{ii) separate Dev and Ops teams with a facilitator(s) in the middle}; \emph{iii) small Ops team with more responsibilities for Dev team}, and \emph{iv) no visible Ops team}. Structures \emph{i} and \emph{iii} map to classical DevOps in our taxonomy. Structure \emph{ii} corresponds to the adoption of a DevOps team as a bridge between development and operations. One of our interviewees reported that such a pattern occurred in the past in his organization (\#I4) and we also observed the DevOps team as a committee bringing together development and operations leadership in another organization (\#I36); therefore, DevOps teams serving as bridges are likely no longer common. Finally, structure \emph{iv} maps to cross-functional teams. Shahin~\emph{et al.} did not identify the platform teams structure, the most promising alternative we found in relation to delivery performance. {\destaque Moreover, their work does not present the notion of transitioning between the structures.}

Shahin~\emph{et al.}~\cite{shahin2017structures} also explored the size of the companies adopting each structure. They found that structure \emph{i} is mainly adopted by large corporations, while structure \emph{iv} was observed mainly in small ones. These findings are corroborated by our data in Table~\ref{tab:structures-and-sizes}: classical DevOps was less observed in small organizations, while cross-functional teams were not prevalent in large organizations. However, other factors may be involved in adopting an organizational structure. Better understanding these factors and their forces are in our plans for future work.

In a recent paper, Shahin and Babar recommend adding operation specialists to the teams~\cite{shahin2020architecture}, which favors cross-functional teams with dedicated infra professionals.

\section{Limitations}
\label{sec:limitations}

The reader must take into account the typical limitations of taxonomy theories. The most important limitation is that they rarely make probabilistic predictions in terms of dependent and independent variables, as variance theories, common in Physics, do~\cite{ralph2019taxonomies}. Thus, it is not proper to discuss, for a grounded theory, the number of interviewees in terms of statistical sampling.

A limitation for credibility and dependability~\cite{guba1981criteria} is the anonymity of our participants, since it is not possible to replicate the interviews with the same people. However, there is a crucial trade-off here: some participants might not accept an invitation to participate in a non-anonymous interview. Moreover, anonymity copes with social desirability bias~\cite{dodou2014desirability}; interviewing the same person after some time does not guarantee the same responses.

Although we relied on previous work~\cite{forsgren2018accelerate} to adopt the delivery performance construct, we needed to adapt the original thresholds due to the reasons exposed in Section~\ref{sec:delivery-performance}. We also did not differentiate low from medium performers, as we judged that such differentiation would not help distinguish how the structures contribute to delivery performance, partly because low and medium performers are much more similar among themselves than when compared to high performers~\cite{puppet2017devops}.
Therefore, by changing some decisions related to delivery performance, another researcher could reach different conclusions based on the same data. We handled this concern mainly by stating our definitions regarding this topic. Nonetheless, GT does not guarantee that two researchers working in parallel with the same data would achieve identical results~\cite{glaser1999gt}.

In an ideal process, different researchers would analyze the interviews independently, avoiding some possible agreement bias favored by reviews. However, the cost for independent analysis would be too onerous (analyzing one interview takes hours). Regarding the review process, instead of relying on the transcripts produced by a single author, all researchers could have listened to the original records. Yet, such an approach also would be too costly in terms of time commitment. Alternatively, two co-authors took part in a few initial interviews to standardize the transcription procedures and assess how the first author conducted these interviews.

We are aware that large companies usually present groups at different maturity levels, and that such groups could be classified differently if we had chosen different interviewees. To verify this, we interviewed two persons from the same company (\#I16 and \#I18) working in different teams. We noted that, indeed, the organizational patterns were not identical. This effect also happens when transitioning from one structure to another, since transitioning can be a long process and take different paces at different organization segments. Therefore, the reader must note that our descriptions characterize our respondents’ contexts, not the organizations in their totality.

Finally, one particular characteristic of software development teams is that they are not fixed; developers often move in and out of a team. Therefore, there is no reason to think about the proposed organizational structures as immutable. Instead, readers should consider our organizational structures when moving toward a continuous delivery scenario. After becoming comfortable and starting to excel with continuous delivery, if needed, practitioners could adapt the organizational structure employed to become effective in other contexts.

\section{Conclusion}
\label{sec:conclusions}

In this paper, we presented an emerging grounded theory addressing the research question ``\emph{What organizational structures do software-producing organizations adopt to \\ structure operations activities (application deployment, infrastructure setup, and service operation in run-time) \\ among development and infrastructure groups?}'' We found four organizational structures:

\begin{enumerate}
\item Siloed departments (with seven core properties);
\item Classical DevOps (with eight core properties and one supplementary property);
\item Cross-functional teams (with four core properties and three supplementary properties);
\item Platform teams (with nine core properties and three supplementary properties).
\end{enumerate}

Beyond the seven supplementary properties directly related to one structure each, we also found two other supplementary properties applicable to more than one structure. The core properties describing each organizational structure plus the supplementary properties answer the sub-research question ``\emph{What are the properties of each of these organizational structures?}'' 

Moreover, this study answers the sub-research question ``\emph{Are some organizational structures more conducive to continuous delivery than others?}'' by raising the emerging hypothesis that the \textbf{platform team} structure is more suitable for achieving continuous delivery given its relationship with delivery performance. We also observed that siloed departments relate more to organizations with less-than-high performance. Additionally, there was no apparent relation between delivery performance and the other structures (classical DevOps and cross-functional teams). Nonetheless, we critically state that reaching high delivery performance is not a need for every software-producing organization, as adopting platform teams is not adequate for every organization. {\destaque A promising topic for future research is investigating whether the organization domain influences the need of high delivery performance.}

Our work has implications for \textbf{practice}. Software companies could realize that there are several kinds of organizational structures they could adopt to excel in continuous delivery and plan which organizational structure they are interested in moving to, maximizing their chances to succeed in the transition. Further, we clarified the roles that the participants have to play in each organizational structure. This evidence could help practitioners cooperate with less friction toward organizational transformation.

Our work also has implications for \textbf{research}. The elements of our taxonomy, in Figure~\ref{fig:taxonomy}, provide a common vocabulary to support the formulation of new research questions. For example, researchers can investigate the impact of each property on other perspectives (e.g., software architecture, security, database management). Another valuable endeavor would be to investigate the relation between our taxonomy's elements and the internal organization of development and operation groups, especially platform teams.

In addition, we observed that test automation is still not adequately practiced in many software companies, as also noted by Olsson~\emph{et al.}~\cite{olsson2012stairway}, and that the lack of tests is a limiting factor for achieving high delivery performance. This suggests research opportunities for proposing automated test generation techniques, especially in environments in which tests are intrinsically difficult, such as games or IoT. Moreover, we noticed participants practicing DevOps while maintaining a monolithic application, which contrasts with the literature that strongly associates DevOps with microservices. This suggests researchers should further investigate the differences in applying DevOps to monoliths and microservices-based systems. Similarly, since we also observed some participants maintaining a monolith core with peripheral microservices and achieving high delivery performance with the microservices, researchers could propose novel techniques and tools that could (semi-) automatically extract peripheral services from monolith applications. 

For future work, we plan to better delineate the forces that drive organizations to choose different structures. We also intend to employ our taxonomy to discuss scenarios of corporations not visited by us yet, which can corroborate the theory's resonance.